\documentclass[10pt,sigconf,letterpaper,screen]{acmart}
\usepackage{graphicx}

\usepackage[english]{babel}
\usepackage{blindtext}

\usepackage{subcaption}

\usepackage{algorithm}
\usepackage{algorithmic }

\usepackage[toc,page]{appendix}

\usepackage{colortbl}
\definecolor{palegreen}{rgb}{0.6, 0.98, 0.6}
\definecolor{White}{gray}{1}
\definecolor{Gray}{gray}{0.9}
\definecolor{black}{gray}{0.6}
\definecolor{Orange}{RGB}{255,153,100}
\definecolor{Bluee}{RGB}{225,250,255}
\definecolor{Blue}{RGB}{185,210,255}
\definecolor{Blu}{RGB}{145,170,255}
\definecolor{Bl}{RGB}{110,135,255}
\definecolor{Green}{RGB}{130,255,130}
\definecolor{Purple}{RGB}{183,130,255}
\definecolor{LightRed}{rgb}{1,0.3,.3}
\definecolor{LightGreen}{rgb}{0.26,1,0.6}
\definecolor{LightBlue}{rgb}{0.26,0.6,1}
\definecolor{ACMLightBlue}{cmyk}{0.49,0.01,0,0}
\newcolumntype{g}{>{\columncolor{Gray}}c}
\newcolumntype{w}{>{\columncolor{Bluee}}c}
\newcolumntype{d}{>{\columncolor{Bluee}}l}
\newcolumntype{b}{>{\columncolor{Blue}}c}
\newcolumntype{r}{>{\columncolor{Blu}}c}

\renewcommand\footnotetextcopyrightpermission[1]{} 
\setcopyright{none}
\setcopyright{acmcopyright}
\setcopyright{acmlicensed}
\setcopyright{rightsretained}
\setcopyright{usgov}
\setcopyright{usgovmixed}
\setcopyright{cagov}
\setcopyright{cagovmixed}

\copyrightyear{2023}
\acmYear{2023}
\setcopyright{acmcopyright}
\acmConference[Preprint]{}{2023}{Canada}
\acmPrice{15.00}
\acmDOI{10.1145/3387514.3405892}
\acmISBN{978-1-4503-7955-7/20/08}

\begin{document}
\title{Internet Congestion Control Benchmarking}

\author{Soheil Abbasloo}


\begin{abstract}
How do we assess a new Internet congestion control (CC) design? How do we compare it with other existing schemes? Under what scenarios and using what network parameters? These are just a handful of simple questions coming up every time a new CC design is going to be evaluated.
Interestingly, the number of specific answers to these questions can be as large as the number of CC designers. 
In this work, we aim to highlight that the network congestion control, as a hot and active research topic, requires a crystal clear set(s) of \textit{CC Benchmarks} to form a common ground for quantitatively comparing and unambiguously assessing the strengths and weaknesses of a design with respect to the existing ones.
As a first step toward that goal, we introduce general benchmarks that can capture the different performance of the existing Internet CC schemes. Using these benchmarks, we rank the Internet CC algorithms and illustrate that there is still lots of room for more innovations and improvements in this topic.
\end{abstract}

\maketitle

\section{Introduction}
\subsection{Setting the Context}
After about four decades of the first series of "congestion collapses"~\cite{collapse} in the Internet and one of the first CC designs~\cite{tahoa}, controlling congestion in the network remained a hot, active, and challenging topic that attracts new designs every year
~\cite{orca,deepcc,copa,bbr2,cubic,newreno,indigo, aurora, sprout, remy, vivace, highspeed, veno,hybla,ledbat,cdg,yeah,xcp,yeah,verus,exll,natcp,c2tcp2,sage}.
One of the main reasons behind the importance of this topic is the fact that CC algorithms directly impact the end-user quality of experience. That is why the improvements of the underlying network technologies, the arrival of new applications, and the changes in the end-users demands always push the community toward introducing \textit{better} CC designs.
Although there is a consensus on the need for novel CC designs in the community, when it comes to \textit{how} to identify better ones and questions such as how we assess a new CC algorithm, how we compare it with other existing designs, under what network scenarios, and using what network parameters, 
there still is a lack of crystal clear consensus among the community. 
In other words, we lack useful \textit{benchmark(s)} in the community to have clear, fair, and universal comparisons between various CC algorithms. 



\subsection{Motivations}
\label{sec:motive}


\textbf{The Benchmarking in Other Communities:}
Using benchmark(s) to assess the performance of a new algorithm (or simply put: benchmarking it) is a very primary and common mechanism that has worked as a cornerstone for driving the innovation in different communities spanning from machine learning (ML) to hardware communities. For example, the ImageNet benchmark~\cite{imagenet,imagenet_,imagenet2,imagenet3} has been widely recognized for boosting innovation in the deep learning context and stimulating competition for improving computer image analysis. 
Among other examples of established simple benchmarks that are used enormously in the ML literature especially in the reinforcement learning topic either for testing the functionality of algorithms or positioning them among existing ones, the HalfCheetah, Hopper, Walker2d, Swimmer, Ant, and Humanoid benchmarks can be named~\cite{openai}.



\noindent\textbf{From the CC Reviewers/Critics' Perspective:}
When reviewing a CC work, usually, one of the main questions that a reviewer faces is whether the new scheme is evaluated in a fair manner compared to the existing CC schemes. 
The general tendency of the designers to show the \textit{good} aspects of their new CC designs can unintentionally impact the scenarios or network parameters they use for highlighting the benefits of their design compared to the existing ones. 
In other words, the issue is about the scope of the evaluations and making sure that the presented results are representative enough and not cherry-picked in the worst case.

\begin{figure}[!t]
    \centering
        \includegraphics[scale=0.4]{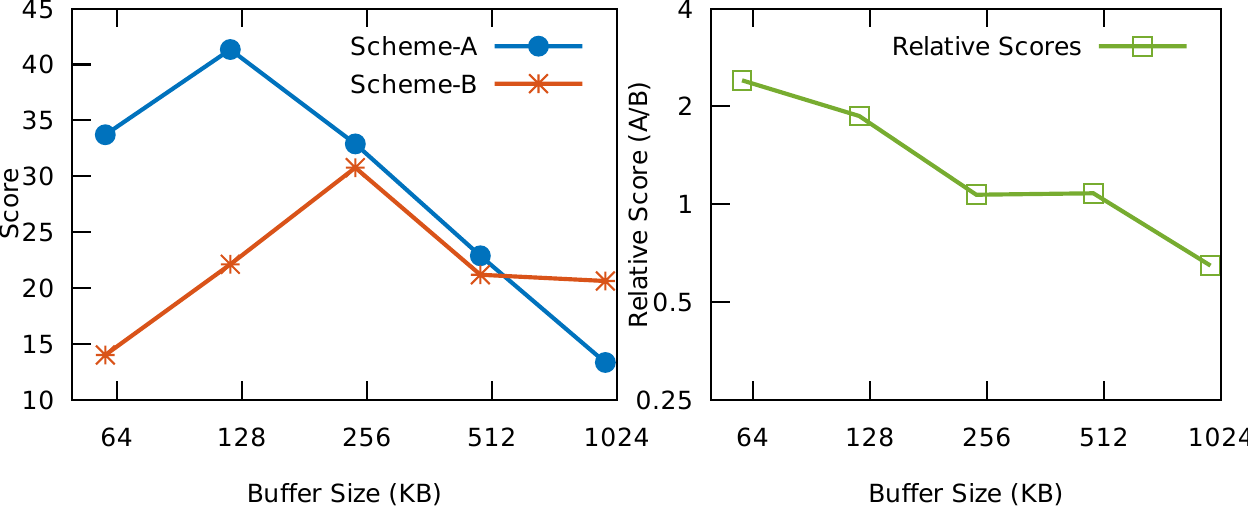}
        \caption{The impact of bottleneck buffer size on the relative performance of two CC schemes. In [64, 256)kB region, scheme A outperforms B, in [256, 512]kB region, both schemes perform roughly similar, and in (512, 1024]kB region, scheme B outperforms A.} 
        \label{fig:score_buffersize}
\end{figure}

\begin{figure}[!t]
    \centering
        \includegraphics[scale=0.4]{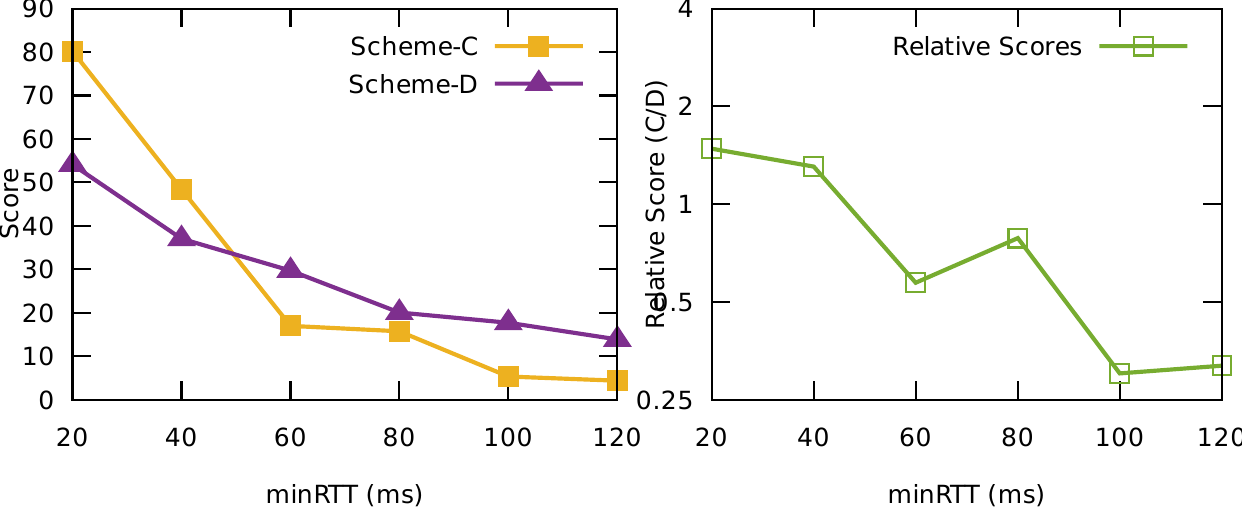}
        \caption{The impact of minimum delay of a network on the relative performance of two CC schemes. In [20, 50)ms region, scheme C performs better than D, while in (50, 120]ms region, scheme D outperforms C.} 
        \label{fig:score_mrtt}
\end{figure}

To show that, we perform two very simple experiments following the settings detailed in Section~\ref{sec:scenarios}. In particular, we set up a simple single bottleneck link where we have control over three main network parameters named: minimum end-to-end delay of the network (minRTT), buffer size, and link capacity (BW). In the first experiment, we choose two Internet CC schemes (Scheme A: TCP Cubic~\cite{cubic} and Scheme B: LEDBAT~\cite{ledbat}) to send and receive traffic through this network and set BW to 48Mbps and minRTT to 40ms. Then, we change the underlying buffer size of the network to capture its impact on the overall performance of these schemes. To that end, we use the score defined in Eq.~\ref{eq:score_p} that encapsulates the gained throughput and delay of a CC scheme (a scheme with a higher score is a better scheme). As Fig.~\ref{fig:score_buffersize} illustrates, it is clear that the relative performance of Schemes A and B depends heavily on the chosen underlying buffer size of the network. So, performing experiments only in one of the three regions mentioned in Fig.~\ref{fig:score_buffersize}, cannot lead to any meaningful comparison of these schemes.
In the second experiment, we use the same network, set BW to 48Mbps, the buffer size to 5$\times$BDP, and change the minRTT value of the network. Here, we use two other Internet CC schemes (Scheme C: TCP Vegas~\cite{vegas} and Scheme D: BBR~\cite{bbr}). Again, the results reported in Fig.~\ref{fig:score_mrtt} illustrate that the relative performance of Schemes C and D depends heavily on the chosen minimum delay of the network. 

The mentioned issues can highlight the fact that using transparent and clear CC benchmark(s) can greatly increase the trust in the reported results of a new design, significantly facilitate the process of reviewing CC algorithms, and avoid reviewers' time and efforts to be spent on the verification of the scenarios, settings, and test methodologies used throughout a work.


\noindent\textbf{From the CC Designers' Perspective:}
Assuming that the first potential critics of an algorithm are the designers of it and they can improve their works by criticizing it before anyone else can, all the benefits discussed earlier can be considered as the benefits of benchmarking during the design of a CC algorithm. Moreover, benchmarking a design can work not only as a vigorous approach for observing the strengths and weaknesses of a design with respect to the existing ones, but also as a mechanism to perform functionality tests of it. In other words, benchmarking can facilitate and accelerate the design process by revealing the problems of a CC scheme during the design process.
In addition, benchmarking can help designers to focus on designing better-performing algorithms rather than spending time on finding test scenarios or corner cases where other schemes perform poorly.

\noindent\textbf{ML and the New Wave of CC Designs:}
The recent advances in machine learning techniques and algorithms have impacted different research communities including the system and network community. In particular, a new wave of CC designs based on (or inspired by) ML has been proposed in the recent years (e.g., ~\cite{orca, aurora, indigo, deepcc, vivace,remy})
and likely in the coming years, there will be more of them out there. 

Generally, learning-based CC schemes are designed and trained over a set of network scenarios and later tested over other scenarios where they are shown to perform very well. Based on these performance reports, the designers may conclude that their scheme poses certain properties. For example, the authors of Indigo mention "We find that Indigo consistently achieves good performance"~\cite{indigo} or PCC's authors say that "PCC achieves consistent high performance"~\cite{pcc}. However, usually, these kinds of statements will be shown to generally not hold true (e.g. check the poor performance of Indigo in~\cite{orca} or the poor performance of PCC in~\cite{copa}).

The main issue here is not with certain schemes or certain claims. The issue is that these general statements are ambiguous and susceptible to multiple interpretations (or simply put they are not comparable), because they are not reported using the same setting or over the same scenarios. That's why usually the reported results of CC schemes create confusion and consequently, make it very difficult to determine whether the improvements reported over the prior state-of-the-art CC designs are meaningful. Benchmarking is a powerful mechanism to prevent such ambiguities.






\subsection{Summary of This Work}
Considering the motivations discussed, our main goal in this work is to highlight 
the need to have and use CC benchmarks. On the one hand, the CC benchmarks can lead to producing verifiable, comparable, and crystal clear statements about the performance of CC designs and make it possible to rank them among the existing approaches, a ranking that can be reproduced and trusted. On the other hand, CC benchmarks can stimulate more competition and greatly accelerate more innovations in this topic.  
As a first step toward that end, in this paper, we 
introduce a set of benchmarks named the CC-Bench1 and the CC-Bench2 for evaluating Internet CC schemes (Section~\ref{sec:benchmarks}). In particular, the CC-Bench1 focuses on the single-flow performance of Internet CC algorithms with respect to their gained throughput and round-trip delay. On the other hand, the CC-Bench2 provides a common ground for assessing the TCP-friendliness criterion of Internet CC schemes. As part of each benchmark, we present the notion of the scores and the winner schemes to facilitate the quantitative comparison of CC schemes.
Using these two benchmarks ({publicly available at~\cite{tcp_bench_repo}}), we examine the performance of twenty two different Internet CC algorithms and rank them based on their gained scores in each benchmark (Section~\ref{sec:eval}). These rankings point to some interesting findings. For example, when considering the CC-Bench1, the top-ranked scheme, that happens to be a ML-based approach, is able to get the best score in only about 33\% of the scenarios! That means in any of the remaining 67\% of the scenarios, the top-ranked scheme performs at least worse than one of the twenty one remaining schemes. This can show the remaining opportunities in this domain and provide a base challenge for observing more innovations and improvements.

\section{Related Work}


There are vast body of work in the literature centered around theoretically analyzing performance of different CC schemes or modeling certain classes of designs such as AIMD (additive increase multiplicative decrease) protocols \cite{macroscopic,reno_model,cc_math}. These models and analysis always can improve our understanding of CC schemes. However, since these analysis and modelings are usually based on simplified assumptions about the underlying networks (e.g., Poisson arrival of traffic, M/M/1 queue models, etc.), they do not suffice and we need to always accompany them with a lot of experiments to really observe the relative performance of CC schemes. 

To perform extensive experiments, from long time ago, network community has adapted to use network simulators and emulators. Frameworks such as NS-2~\cite{ns2}, NS-3~\cite{ns3}, NetEm~\cite{netem}, Mininet~\cite{mininet}, and Mahimahi~\cite{mahi} provide different underlying parameters and components that can simulate/emulate various network models/functionalities. 
However, the specific choice of all these various parameters and settings, which as shown in Section~\ref{sec:motive} will impact the output of evaluating network protocols and designs such as CC schemes, remains open. In this work, we exploit these tools, in particular Mahimahi, and take a pragmatic approach by introducing a set of benchmarks that do not delegate the choice of \textit{proper} settings for evaluating CC schemes to the CC designers themselves. 

The goal of evaluating CC schemes over the same network scenarios is not necessarily new. As a recent important example aiming this problem, Pantheon~\cite{indigo} can be named. Pantheon hosted a testbed of measurement nodes on the Internet and performed periodic experiments with some of CC designs and generated periodic reports on relative performance of the tested schemes. However, it is for more than three years that the platform is not maintained anymore. Inspired by these efforts, we argue that having CC benchamrks that can be executed locally can have their own advantages over real testbeds. First, local CC benchamrking can prevent the costs and fees associated with maintaining a centralized platform and hosting a common testbed. Second, locally available CC benchmarks can potentially cover more variety of scenarios compared to real testbeds. Third, there is no control over the \textit{other} competing flows in a real testbed, so lots of data are required to make sure that on average results indicate something meaningful. Also, it will be hard to assess properties such as TCP-friendliness of a CC scheme over an uncontrolled network. 






\section{CC Benchmarking}
\label{sec:benchmarks}
The process of benchmarking usually consists of two main elements: (1) the metrics and (2) the set of environments, tasks, or scenarios. So, here, we start with these components and elaborate on the details of each element in our benchmarks.
\subsection{The Metrics \& the Notion of Scores}
When discussing the benchmarking of algorithms/systems, the very first question that comes to mind is what metric we should use for comparing them quantitatively in a way that it'll be transparent and clear enough to reveal the differences among various systems. This question brings up the need for having clear quantitative metrics that can represent the main objective(s) in the targeted domain. 

\noindent\textbf{The Right Metric vs. a Good Metric:} If we consider the CC problem as a subset of general decision-making problems, wearing an Operations Research glass may lead us toward formulating the CC problem as a classic optimization problem where the key objective is to maximize a given utility function that represents the problem. However, when it comes to how to model a CC problem and choose the right utility function, which should represent the main objective in the context of the CC problem, there are no unique answers out there. An interesting sign of that fact can be found among the recent learning-based CC designs. By looking at the details of these works, one can realize that utility functions used in these works span a wide range of possible candidates from more classic ones such as the Power~\cite{power} (or modified versions of that as in~\cite{orca, remy}) to very handcrafted and engineered functions including the first (or higher) derivatives of delay, throughput, etc. (e.g. as in~\cite{vivace}). That's why here, we do not try to formulate the Internet CC problem or come up with \textit{the} right utility function. Instead, we wear a System Designer glass, and look for \textit{a good} metrics that can practically encapsulate the general properties of a good CC scheme from the end-user perspective. 


The early practical CC schemes were designed centered around maximizing users' throughput while maintaining fairness among different flows. This main practical goal led to an enormous number of heuristic AIMD algorithms for Internet CC (e.g.~\cite{tahoa,reno,newreno,bic,cubic}), and even today, the default CC algorithm in most platforms is still an AIMD algorithm that had the same intrinsic objective during the design (TCP Cubic~\cite{cubic}). The throughput-oriented nature of CC designs was a practical and acceptable assumption mainly due to the throughput-oriented nature of the dominant applications at the time. Recently though, the emerging applications (such as AR/VR, online gaming, tactile Internet, vehicle-to-vehicle communications, etc.) and their intrinsic delay-sensitive natures have spurred a new wave of CC designs that try not only to maximize user's throughput, but also to reduce the delay~\cite{bbr,bbr2,orca,vivace,copa,deepcc,sprout}.

In addition, the notion of incremental growth of the Internet over existing technologies and protocols has brought up another key practical challenge for designing new Internet CC schemes named TCP-friendliness. The challenge of TCP-friendliness comes from the fact that a new Internet CC scheme should not only provide a good throughput and delay performance, but also should be able to compete fairly with the default and established Internet CC scheme used by the majority of devices on the Internet. A scheme that is very aggressive toward the default scheme usually loses ground among the community and a scheme that is very polite/shy usually is not considered a practical solution. 

So, putting it all together, we can reason about the performance of a CC scheme when we consider its throughput, delay, and TCP-friendliness. That said, to capture these three main metrics in different scenarios, we define two different scores for a flow $F_c$ that uses the CC scheme $c$.
To reflect the throughput and delay performance metrics of $F_c$, we use a modified version of Power~\cite{power} and define $S^c_p$ for a flow $F_c$:
   \begin{equation}
   \footnotesize
   \label{eq:score_p}
       S^c_{p} = \frac{r_c^\alpha}{d_c}
   \end{equation}
   where $r_c$ and $d_c$ are the average delivery rate and the average round-trip delay of $F_c$, respectively, and $\alpha$ is a coefficient determining the relative importance of throughput and delay ({unless otherwise mentioned, we set $\alpha=2$}). A bigger value of $S^c_p$ indicates better throughput and delay performance for the CC scheme $c$. 
To reflect the TCP-friendliness metric, in a multi-flow scenario, we define $S^c_{fr}$ as: 
   \begin{equation}
   \footnotesize
   \label{eq:score_fr}
       S^c_{fr} = |f_c-r_c|
   \end{equation}
   where $f_c$ is the expected average fair share of $F_c$, when competing with the flows with default CC scheme, and $r_c$ is the actual achieved average delivery rate of $F_c$. A smaller value of $S^c_{fr}$ indicates a better TCP-friendliness property for $c$.


\subsection{Scenarios Covered}
\label{sec:scenarios}
On the one hand, clearly, a benchmark cannot cover all possible scenarios and putting too many scenarios may lead to some practical issues such as the resources required to run the benchmark.
On the other hand, a benchmark that spans only a few cases will not be able to effectively reveal differences among various CC schemes. 
So, as a general rule of thumb and as a first step to making CC benchmarks, we tried to keep the set big enough to see differences among the CC schemes and small enough to make them practical benchmarks. One can imagine that in the future more sets of scenarios can be added to the current list (as is the case in other communities~\cite{openai}).

In this section, we elaborate on the network model used in the benchmarks and describe the range of underlying network parameters and the details of the scenarios.

\noindent\textbf{Network Model:} To keep things simple, we use a single bottleneck link model of the network. This model can be specified using three main parameters: (1) bottleneck link capacity (BW), (2) minimum end-to-end delay (minRTT), and (3) bottleneck buffer size ($qs$). We will show later in Section~\ref{sec:eval} that this simple model is good enough to reveal the differences among various Internet CC schemes. 
To emulate this network model, we use Mahimahi~\cite{mahi} that creates TUN/TAP interfaces on Linux OS and provides us with control over the BW, minRTT, and $qs$ values. In addition, we optimize different Linux Kernel parameters (in particular, TCP related ones) to have best performance~\cite{tcp_bench_repo}.

\noindent\textbf{Range of Underlying Network Parameters:} 
For choosing the range of parameters, we considered typical Internet scenarios with a focus on normal end users on the Internet and some limitations of Mahimahi such as its large overhead for the large values of BW. That said, the benchmarks cover BWs from a few Mbps to about 200Mbps, minRTTs from 10ms to 160ms, and $qs$ from $\frac{1}{2}\times$BDP to $16\times$BDP\footnote{The problem of how to set buffer sizes in a network is for itself an interesting (and not a simple) problem that still attracts new solutions. E.g. check out~\cite{buffersize-workshop} and the papers therein for a recent workshop dedicated to this issue.}. 

We categorize benchmarks into two separate groups called CC-Bench1 and CC-Bench2 which include single-flow and multi-flow scenarios, respectively. 
\subsubsection{The CC-Bench1}
This set of benchmarks consists of single flow scenarios where schemes are evaluated with respect to $S^c_p$ score that reflects their throughput and delay performance. The CC-Bench1 includes two main classes of scenarios: (1) the flat scenarios and (2) the step scenarios. 

\noindent\textbf{The Flat Scenarios:} This set of scenarios represents general wired scenarios on the Internet. As its name suggests, it includes wired links with constant/flat bandwidths throughout the experiments. The ranges of BW, minRTT, and $qs$ are $[12, 192]$Mbps, $[10,160]$ms, $[\frac{1}{2},16]\times$BDP, respectively.

\noindent\textbf{The Step Scenarios:} The flat scenarios alone cannot show the performance of CC schemes over a more dynamic network. So to answer questions such as how a CC scheme reacts when BW reduces or increases suddenly, we bring up the step scenarios. In these scenarios, we start with a given network BW ($BW_1$) and after a specific period of time, we change the underlying BW of the network to $m\times{BW_1}$. Then, we repeat this cycle until the end of the evaluation. The $m$ value is chosen from $(0.25,0.5,2,4)$ list. 

There are a couple of points here. First, we observed that some of the CC schemes perform periodic tasks (e.g. every 10s, BBR reduces its congestion windows to a few packets~\cite{bbr} to observe the change of minRTT). So, to keep the benchmarking process fair, we tried to avoid having overlaps between the times of changes and these periodic tasks by choosing a safe time period for the changes. That said, we perform the changes every 7 seconds\footnote{As far as we know, this value does not overlap with any settings for the existing CC schemes.}, unless mentioned otherwise. Second, we observed that for large values of BW, Mahimahi's overhead increases to a point that it tangibly impacts the results. To prevent these unwanted impacts, when changing BW, we always choose to be under 200Mbps. That means if $BW_1$ is 96Mbps, we choose $m<4$.
The range of other parameters is similar to flat scenarios.
For both flat and step scenarios, each CC scheme sends traffic for 30s.

\subsubsection{The CC-Bench2}
This set of benchmarks provides scenarios for assessing the TCP-friendliness of Internet CC schemes. To that end, we let TCP Cubic, which is the default CC scheme in most platforms (including Linux, Windows, and macOS), compete with the CC scheme under the test for accessing a shared bottleneck link and we capture the performance of schemes with respect to $S^c_{fr}$ score. Similar to the CC-Bench1, the three main network parameters are changed to make different scenarios. The ranges of minRTT and BW values are similar to CC-Bench1. In addition, we at least let the bottleneck link have $1\times$BDP buffer size to be able to effectively absorb more than one flow during the tests. In particular, we choose $qs$ from $[1,16]\times$BDP range.

In a general Internet scenario, with a good probability, we can assume that a new incoming flow will observe flows controlled by the default CC scheme on the bottleneck link. This comes from the definition of a default CC scheme and its property of being used by the majority of flows. Therefore, we let TCP Cubic come to the network earlier than the CC scheme under the test.
When buffer size increases, generally, it takes more time for flows to reach the steady-state (if any). In our experiments, we observed that reaching a fair-share point may take more than a minute (even when both flows are Cubic flows). So, in the CC-Bench2, we let flows send their packets for 120s to make sure that the results can present meaningful TCP-friendliness scores.

\subsection{The Notion of the Winner Schemes}
A more classic way of looking at who should be called the winner in a certain scenario may lead us toward recognizing the CC scheme with the best score gained throughout a scenario as the winner in that scenario. However, there are two issues with this way of identifying a winner. 

First, since the scores defined in equations~\ref{eq:score_p} and ~\ref{eq:score_fr} are Real numbers, their absolute values can differ slightly for two CC schemes. So, if we simply perform a mathematical comparison between scores, these slight differences can impact the choice of the winner in a scenario. That said, instead of picking the CC scheme with the best score as the winner, we pick all CC schemes with scores less than 10\% worse than the best score as the winners of a scenario. In other words, any scheme with at most 10\% lower performance than the best performing scheme is included in the winner list of that scenario. 

Second, simply assigning a number to the performance of a scheme over an entire scenario 
and then comparing these numbers together to decide the winners may smooth out the important differences among the CC schemes. For instance, how fast CC schemes can react to a sudden change in the network may not be visible in an overall score of the scheme over a longer period. To address this issue, we calculate the score of a scheme in separate intervals throughout the experiments and instead of one score, assign four scores corresponding to the performance of the scheme in four different intervals throughout the test. Now, comparing the scores of a certain interval for all schemes can get a better sense of the performance of different schemes. 

Putting all together, we compare the scores of all CC algorithms over a certain interval of a certain scenario and pick the best performing schemes (considering the 10\% winning margins) as the winners. Then, we sweep over all intervals and scenarios. 

\section{The Ranking of the Internet CC}
\label{sec:eval}
In this section, we rank the Internet CC schemes based on their performance in the CC-Bench1 and CC-Bench2 benchmarks\footnote{All benchmarks are run over 48-Core, 256GB RAM servers.}. The non-exhausting list of schemes consists of 22 different Internet CC algorithms. These algorithms include 14 available TCP schemes in Linux Kernel named Cubic~\cite{cubic}, Vegas~\cite{vegas}, YeAH~\cite{yeah}, BBRv2~\cite{bbr2}, NewReno~\cite{newreno}, Illinois~\cite{illi}, Westwood~\cite{west}, Veno~\cite{veno}, HighSpeed~\cite{highspeed}, CDG~\cite{cdg}, HTCP~\cite{htcp}, BIC~\cite{bic}, C2TCP~\cite{c2tcp}, and Hybla~\cite{hybla}, 5 recent ML-Based CC schemes named Orca~\cite{orca}, Indigo~\cite{indigo}, Aurora~\cite{aurora}, PCC Vivace~\cite{vivace}, and DeepCC~\cite{deepcc}, and 3 other CC schemes named Copa~\cite{copa}, LEDBAT~\cite{ledbat}, and Sprout~\cite{sprout}. 

\begin{figure}[!t]
   \includegraphics[width=1.1\linewidth,height=2.5in]
   {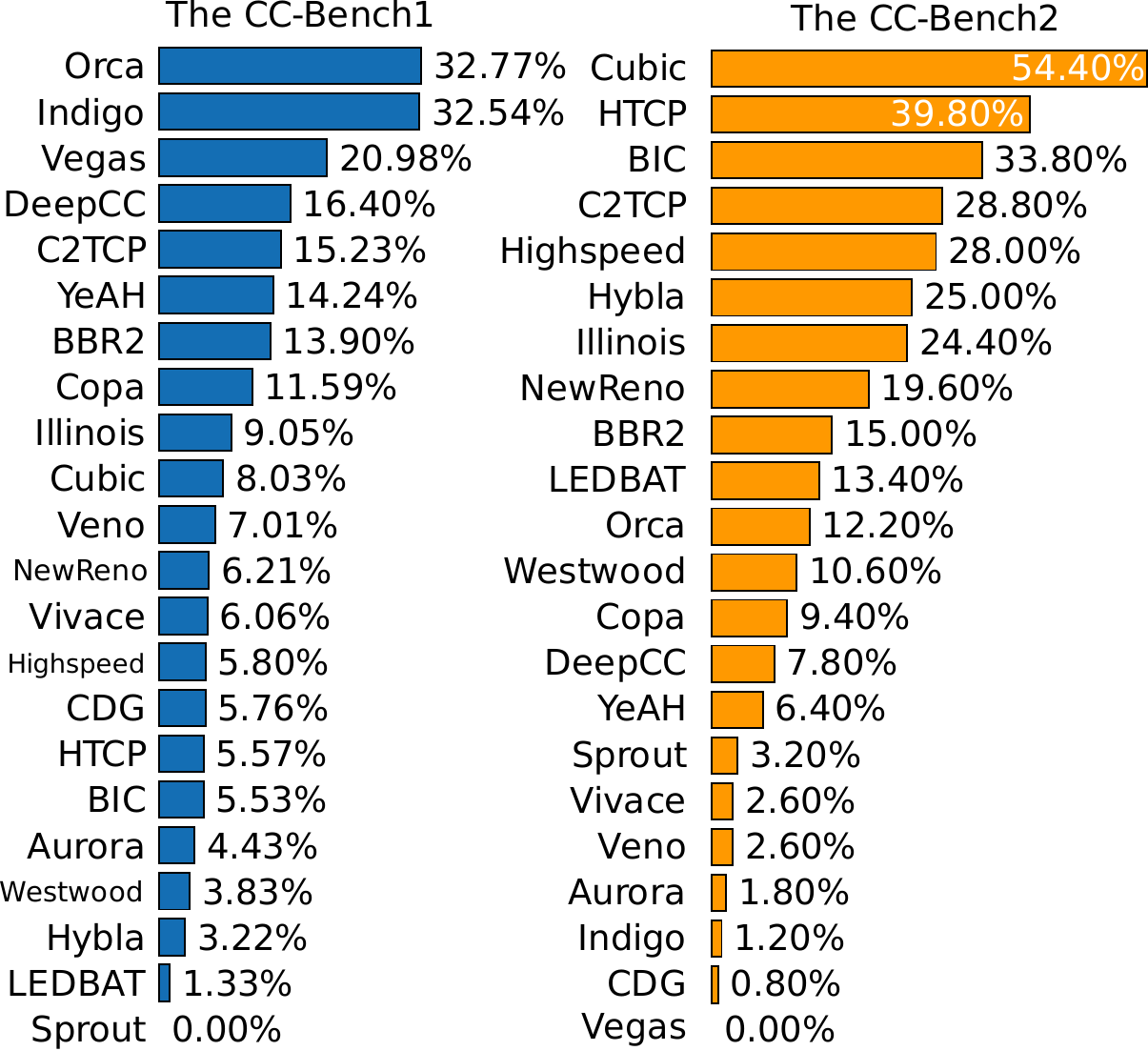}
   \caption{The ranking of Internet CC schemes based on their winning rates in the CC-Bench1 (left) and the CC-Bench2 (right) benchmarks} 
   \label{fig:ranking}
\end{figure}

To report the results, we use the winning rate index defined as the percentage of the times that a CC algorithm was a winner in a benchmark. The rankings of the Internet CC schemes based on their winning rate index for the CC-Bench1 and CC-Bench2 benchmarks are depicted in Fig.~\ref{fig:ranking}. 

\noindent\textbf{Benchmarking to Reveal Opportunities:} 
The very first thing that Fig.~\ref{fig:ranking} illustrates is that none of these Internet CC schemes are perfect and can achieve a 100\% winning rate! A simple CC benchmark such as the CC-Bench1 can reveal the huge opportunities existing in this topic, especially when we consider that the top-ranked schemes in CC-Bench1 (Orca \& Indigo) can be winners in only 1/3rd of the scenarios. How to boost performance further and design a CC algorithm that can achieve higher winning rates is an interesting challenge embracing more innovative designs.  

\noindent\textbf{The Good and the Bad, but not the Ugly:} 
When we look at the results of Fig.~\ref{fig:ranking} from another angle, we may see that having clear CC benchmarks provides us with both the \textit{full-half} of the glass (the part that the designers of the top-ranked schemes can be happy about) and the \textit{empty-half} of the glass (the extent of the remaining performance issues which embrace more novel designs). We think that this is a much more transparent and fair evaluation mechanism compared to when designers of a new CC scheme choose the scope and the details of scenarios in their evaluations.

\noindent\textbf{Getting Inspired by Looking at the Bigger Picture:} 
Results of Fig.~\ref{fig:ranking} indicate a common pattern among the CC algorithms: generally, delay-based schemes are among the top-ranked schemes in CC-Bench1, while throughput-oriented schemes are among the top-ranked schemes in CC-Bench2. This is based on the fact that being TCP-friendly when coexisting with a loss-based scheme and (at the same time) being able to minimize delay and maximize throughput when not competing with a loss-based scheme is very hard and challenging. Interestingly, some of the results in Fig.~\ref{fig:ranking} may point to some pragmatic ways to achieve a better trade-off between TCP-friendliness and throughput/delay performance of a CC scheme\footnote{In particular, C2TCP, BBR2, and Orca seem to have managed this trade-off to some different degrees.}. This notion of potentially getting inspired by the good/bad results of existing schemes is among the benefits that CC benchmarks can bring to the table.


\noindent\textbf{Shedding Light on Neglected Evaluation Aspects:} As expected, Cubic is the top-ranked scheme in CC-Bench2. However, the fact that it cannot achieve a 100\% winning rate may seem to be somewhat unexpected. One reason for this performance lies in a key but usually neglected aspect of CC schemes: the convergence speed of a CC scheme. In particular, although it can be shown that Cubic can eventually approach its fair share when competing with another Cubic flow, its convergence time for reaching that point is not necessarily the best, especially when compared with other top 5 schemes in large $qs$ scenarios\footnote{The mechanism of segmenting an entire run to different intervals and assigning scores to each interval enables us to capture this behavior.}. The bottom line is that CC benchmarks (e.g., CC-Bench2) can potentially shed light on some neglected evaluation aspects of CC algorithms. 






\section{A Brief Discussion \& Final Note}
\textbf{I Think I Can Define Better CC Benchmarks!}
The main point of this work is not to introduce \textit{the} best set of benchmarks, but to highlight the benefits of having CC benchmarks and the need for using them. The CC-Bench1 and the CC-Bench2 benchmarks are by no means the only or the best benchmarks out there. We think that when a community starts using benchmarks, it will not be hard to imagine that as part of an evolutionary process, more attractive benchmarks will be introduced and adopted in the future. 


\noindent\textbf{Are These Benchmarks Enough?} Clearly, one or two benchmarks cannot cover all sets of evaluations and potentially highlight all the specific characteristics of different CC algorithms. That said, the main role of CC benchmarking is not to replace existing detailed stress tests of schemes, but to provide a base and a common ground for unambiguously comparing general results of different works in this domain.

\noindent\textbf{A Couple of Points on the Reported Rankings:}
First, none of the scenarios in our benchmarks were designed to push certain schemes either to the top or bottom parts of the rankings reported. Second, our focus on this work was not to pinpoint the issues with certain CC schemes in certain scenarios. This alone can manifest itself as a good motivation for separate future work.




\noindent\textbf{\underline{Final Note:}}
Although benchmarking is a common practice in different communities, when it comes to CC design, this powerful tool is still missing. With the recent rise of a new wave of CC designs, the need for having such common foundations to unambiguously evaluating new algorithms increases. We hope this work will highlight this need and provide the community with a few samples to continue with. 

\bibliographystyle{ACM-Reference-Format}
\bibliography{reference}

\end{document}